# Z Anomalous Couplings and the Polarization Asymmetry in $\gamma e \to Ze$ [*]

THOMAS G. RIZZO

*Stanford Linear Accelerator Center*

*Stanford University, Stanford, CA 94309*

## Abstract

We extend our previous analysis of the sensitivity of the energy dependence of the polarization asymmetry in $\gamma e \to W\nu$ to the possible existence of anomalous trilinear gauge couplings to the $\gamma e \to Ze$ case. We find that by combining the constraints imposed by both the energy dependence of the total cross section and polarization asymmetry, strong limits on the anomalous $\gamma ZZ$ couplings are obtainable at the Next Linear Collider. Further constraints obtained from a consideration of the angular distribution are briefly discussed.

Submitted to Physical Review **D**.

---

[*]Work supported by the Department of Energy, contract DE-AC03-76SF00515.

# 1  Introduction

The discovery and mass determination of the top quark at the Tevatron[1, 2], combined with the high precision measurements made by the four LEP experiments and SLD at the Z-pole[3] as well as the new $W$ mass value from CDF[4], has shown us that the Standard Model(SM) generally provides a very good description of physics below the electroweak scale. However, there is now some reason to believe that new physics of some kind may be just around the corner. The $R_b - R_c - \alpha_s$ 'crisis' has inspired many[5] authors to contemplate a wide assortment of new physics scenarios as has the apparent observation of an excess of high-$E_T$ jet production by the CDF collaboration[6, 7]. Apart from these 'crises', there are many reasons to believe that some kind of new physics must exist and may not be far away, although how it will first be observed remains unclear.

One of the most sensitive probes of new physics beyond the SM is provided by the trilinear couplings of the gauge bosons[8]. Deviations from the canonical SM values in the form of anomalous couplings, beyond those induced by conventional radiative corrections, are expected to be small, of order $10^{-2}$ or less, on rather general grounds, but may potentially be observable in the future at both hadron and $e^+e^-$ colliders. However, present direct experimental probes of these trilinear couplings are still at a rather early stage–still several orders of magnitude away from theoretical expectations due to various kinds of new physics. Of course the upgrade to the Main Injector at the Tevatron and the advent of the LHC and Next Linear Collider(NLC) will eventually open up interesting ranges of anomalous coupling parameters for exploration and potential new physics discovery.

In our previous work[9], we examined the possibility of using the energy dependence of the polarization asymmetry to probe for non-zero anomalous $\gamma WW$ couplings in the process $\gamma e \to W\nu$ using backscattered laser beams at the NLC. Amongst others, advantages to this process are that it allows us to isolate the $\gamma WW$ vertex, unlike the more familiar $e^+e^- \to W^+W^-$ process where both $(Z,\gamma)WW$ anomalous couplings may participate, and that we can polarize *both* the initial $e^-$ and photon beams. Our analysis showed that indeed the energy dependence of the polarization asymmetry could be sufficiently well determined at the NLC to be useful as a probe of phenomenologically interesting values of the CP-conserving anomalous coupling parameters $\Delta\kappa$ and $\lambda$. This provided an additional weapon in a large arsenal with which such couplings will eventually be determined or constrained.

Unlike the $(Z,\gamma)WW$ cases, the anomalous couplings of the type $\gamma ZZ$ and $\gamma\gamma Z$ have not received very much attention in the literature[8]. In this paper we extend our previous $W$ analysis to the energy dependence of the polarization asymmetry in the process $\gamma e \to Z e$ which probes just these anomalous vertices[10]. *Unlike* the $\gamma e \to W\nu$ case, however, the $\gamma e \to Z e$ process does not isolate either the $\gamma ZZ$ or $\gamma\gamma Z$ set of anomalous $Z$ couplings. In the analysis below we will assume that *only* the CP-conserving, $\gamma ZZ$ anomalous couplings can be non-zero for purposes of simplicity, but we can of course repeat the entire analysis employing instead the $\gamma\gamma Z$ couplings. (We note in passing that additional constraints on these anomalous $\gamma ZZ$ couplings can be obtained through an examination of the $Z$ boson's angular distribution and polarization. These additional measurements will allow for tighter constraints from NLC data than what we obtain below using just the energy dependencies of the cross section and polarization asymmetry. The angular dependence as an additional constraint is briefly discussed in the Appendix.) There are, however, many similarities between these two $\gamma e$ processes. In particular, in both cases the logarithmic integral over the polarization asymmetric part of the cross section that appears in the Drell-Hearn-Gerasimov(DHG) sum rule[11] vanishes[12], if and only if canonical couplings (*i.e.*, gauge couplings) are present, thus implying the existence of a zero in the polarization asymmetry at some value of $\sqrt{s_{\gamma e}}$[9]. (We will sometimes we refer to this polarization asymmetry as the DHG asymmetry, $A_{DHG}$, in our discussion below.) As in the $\gamma e \to W\nu$ case, we will see below in the $\gamma e \to Ze$ case that the number of such zeros is not altered when anomalous couplings are present and that the location of the single zero *is* indeed modified. In the $\gamma e \to Ze$ case another difference with the previously studied $\gamma e \to W\nu$ process arises due to the $u$-channel colinear singularity which is present when the electron mass is neglected. In practice, however, there is a minimum angle (or minimum $p_t$) below which the electron cannot be detected, so that when an experimental cut is required to identify the process in question, the singularity is of course removed. Interestingly, however, the value chosen for this minimum angle($\theta_0$) leads to a modification of the position of the asymmetry zero in terms of the center of mass energy, even in the SM! As we will see from the full cross section expressions below, we can analytically determine the position of the zero ($\sqrt{s_0}$) in the SM for different values of $\theta_0$; we obtain

$$\sqrt{s_0} = 2 \left[ \frac{log\frac{(1+z_0)}{(1-z_0)} + z_0}{2log\frac{(1+z_0)}{(1-z_0)} - z_0} \right]^{1/2} M_Z , \qquad (1)$$

where $z_0 = cos(\theta_0)$ and $M_Z = 91.1884$ GeV is the $Z$ mass[3]. This result is shown numerically in Fig.1. For all reasonable values of $\theta_0$ the zero is seen to arise at rather low energies in comparison to the $\gamma e \to W\nu$ case where the corresponding asymmetry zero occurs near 254 GeV in the SM. Since this zero could have been located anywhere, we are again reasonably lucky that it occurs at energies which will be easily accessible at the NLC. (In fact, the zero is not far from presently available collider energies!) Also, as in the $\gamma e \to W\nu$ case, the asymmetry zero in $\gamma e \to Ze$ occurs not far from the unpolarized cross section maximum, so that we gain the benefit of high statistics.

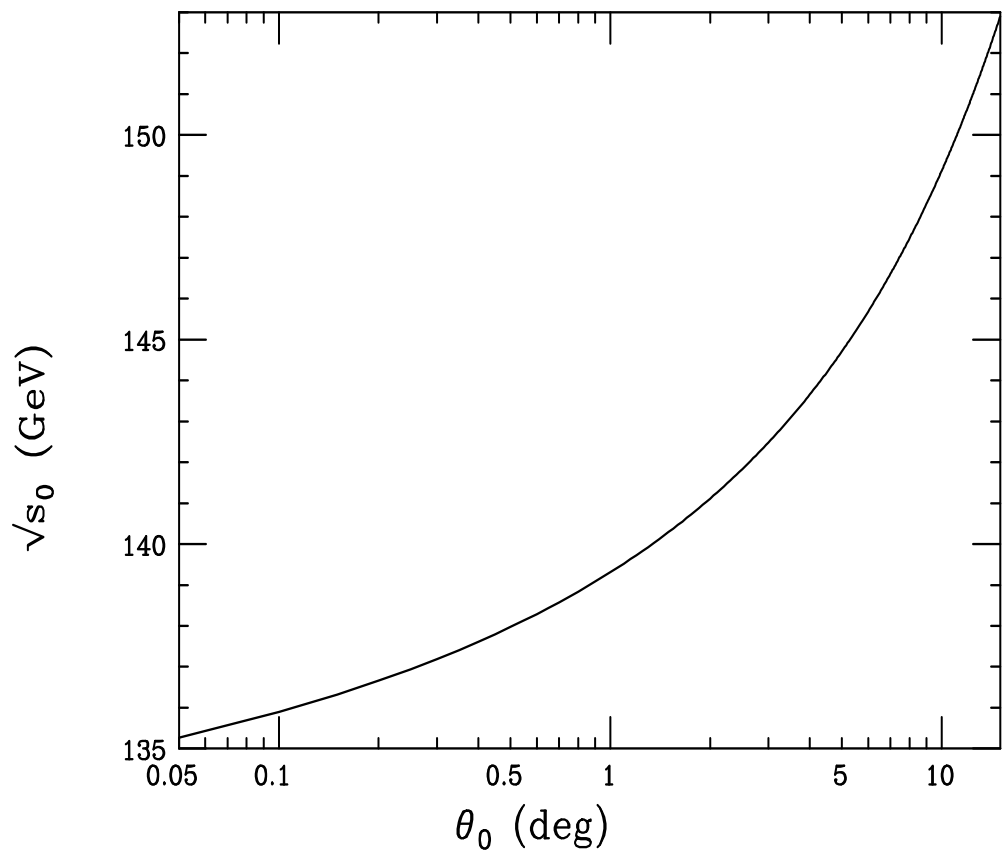

Figure 1: Position of the SM polarization asymmetry zero in $\gamma e \to Ze$ as a function of the angular cut-off $\theta_0$.

## 2 Cross Section and Polarization Asymmetry

To begin this discussion, we first remind the reader how the CP-conserving $\gamma ZZ$ anomalous couplings are traditionally defined. Consider the vertex responsible for the process $Z_\mu(P) \to Z_\alpha(q_1)\gamma_\beta(q_2)$ with the momenta $P$ incoming and both $q_{1,2}$ outgoing. Then the anomalous $\gamma ZZ$ couplings are defined via

$$\Gamma^{\alpha\beta\mu}_{\gamma ZZ} = ie\frac{P^2 - q_1^2}{M_Z^2}\left[h_3^Z \epsilon^{\mu\alpha\beta\rho}q_{2\rho} + \frac{h_4^Z}{M_Z^2}P^\alpha \epsilon^{\mu\beta\rho\sigma}P_\rho q_{2\sigma}\right], \qquad (2)$$

which vanishes if both $Z$'s are on-shell due to Bose symmetry. Note that $h_3^Z$ is associated with a dimension-6 operator while $h_4^Z$ is associated with one of dimension-8. Of course the anomalous couplings $h_{3,4}^Z$ must actually be form-factors which are needed to restore unitarity at high energies and they are thus conventionally parameterized[8] as

$$h_i^Z = \frac{h_i^0}{(1 + q^2/\Lambda^2)^{n_i}}, \qquad (3)$$

where $\Lambda$ is some large scale. In order to compare our results with those obtained previously[8], we choose this scale to be $\Lambda$=1.5 TeV and take $n_{3,4} = 3, 4$ in our numerical analysis below. Keep in mind that the current limits on these anomalous couplings are still rather poor, *i.e.*, $|h_3^0| < 1.6$ and $|h_4^0| < 0.4$, respectively, for the $\gamma ZZ$ case. These limits are several orders of magnitude away from the parameter region we might expect these anomalous couplings to occupy from new physics sources.

With this notation the polarization-dependent $\gamma e \to Ze$ differential cross section can be written as

$$\frac{d\sigma}{dz} = (1 + A_0 P)\sigma_{un} + \xi(P + A_0)\sigma_{pol}, \qquad (4)$$

where $z = cos\theta$ with $\theta$ being the $e^-$ scattering angle, $P$ is the polarization of the initial $e^-$ beam (note that $P > 0$ is left-handed in our notation), $\xi$ is the Stoke's parameter describing the circular polarization of the back-scattered laser photons, and $A_0 = 2v_e a_e/(v_e^2 + a_e^2)$ with $v_e, a_e$ being the usual vector and axial vector couplings of electrons to the $Z$: $v_e = -\frac{1}{2} + 2sin^2\theta_W$ and $a_e = -\frac{1}{2}$. (For numerical purposes we take $M_Z = 91.1884$ GeV, $sin^2\theta_W = 0.2315$, and $\alpha^{-1} = 128.896$.) Defining the overall cross section normalization by

$$\tilde{\sigma} = \frac{\alpha G_F M_Z^2}{4\sqrt{2}s}(1 - \frac{M_Z^2}{s})(v_e^2 + a_e^2), \tag{5}$$

where $\sqrt{s}$ is the $\gamma - e$ center of mass energy, we can decompose the individual polarization (in)dependent parts of the tree level cross section in the limit of zero electron mass as

$$\sigma_{un} = \tilde{\sigma}[a_1 + a_3(h_3^Z)^2 + a_5(h_4^Z)^2 + a_7 h_3^Z h_4^Z],$$
$$\sigma_{pol} = \tilde{\sigma}[a_2 + a_4(h_3^Z)^2 + a_6(h_4^Z)^2 + a_8 h_3^Z h_4^Z], \tag{6}$$

and the $a_i$ are given by

$$a_1 = -\left[(s - M_Z^2)^2 + (u - M_Z^2)^2\right]/us$$

$$a_2 = \left[s(s - 2M_Z^2) - u(u - 2M_Z^2)\right]/us$$

$$a_3 = 4us M_Z^4 \left[(u+s)(u^2 + s^2) - M_Z^2(s^2 + u^2 + 4us)\right]/D$$

$$a_4 = 4us M_Z^4 \left[(u^2 - s^2)(u + s - M_Z^2)\right]/D$$

$$a_5 = us \left[u^5 + s^5 + 3su(u^3 + s^3) + 4s^2u^2(u+s) - M_Z^2[u^4 + s^4 + 2su(u^2 + s^2) + 2s^2u^2]\right]/D$$

$$a_6 = us \left[u^5 - s^5 + 3su(u^3 - s^3) + 2s^2u^2(u-s) + M_Z^2[s^4 - u^4 + 2us(s^2 - u^2)]\right]/D$$

$$a_7 = 4us M_Z^2 \left[s^4 + u^4 + 2su(u^2 + s^2) + 2s^2u^2 - M_Z^2[s^3 + u^3 + us(u + s)]\right]/D$$

$$a_8 = 4us M_Z^2 \left[u^4 - s^4 + 2su(u^2 - s^2) + M_Z^2[s^3 - u^3 + su(s - u)]\right]/D, \tag{7}$$

where $D = 32usM_Z^{10}$ with $u = \frac{-1}{2}s(1+z)$. Note the symmetry of this result under the interchange $h_{3,4}^Z \leftrightarrow -h_{3,4}^Z$. It is also interesting to note that all of the even $a_i$ (and hence $\sigma_{pol}$ itself) are odd under the interchange $s \leftrightarrow u$, while all of the odd $a_i$ are even under this interchange; independently of the existence of the anomalous $\gamma ZZ$ couplings. In addition we see that $a_{4,6,8}$ vanish when $u = -(s - M_Z^2)$ [which corresponds to the point $z = 1 - 2M_Z^2/s$ in the angular distribution], whereas $a_2$ is seen to vanish when $u = -s + 2M_Z^2$ [which corresponds to $z = 1 - 4M_Z^2/s$, and hence lies outside the physical region unless $s \geq 2M_Z^2$]. The terms proportional to $a_3 - a_8$ vanish when $u = -s$ due to the overall factor of $t - M_Z^2$ in the definition of the $\gamma ZZ$ anomalous couplings. Similarly, the terms proportional to $a_3 - a_8$ are also seen *not* to exhibit the $u$-channel pole structure of the SM pieces. The total cross section is directly obtained by integration of the above differential expression subject to a suitable angular cut as discussed above.

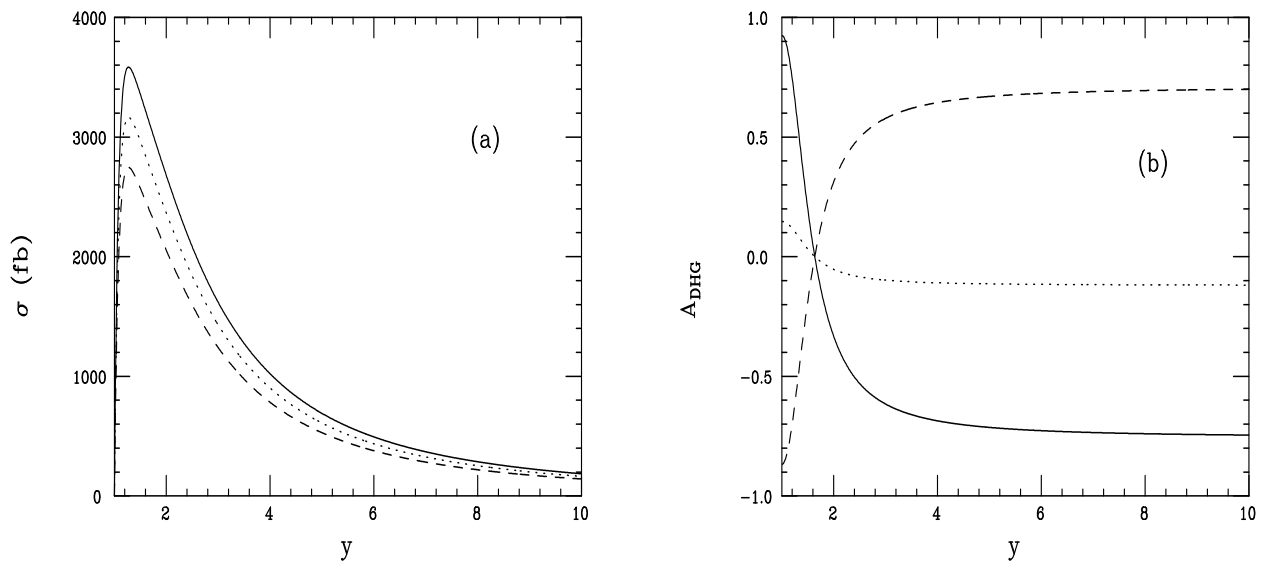

Figure 2: (a) Cross section and (b) polarization asymmetry in $\gamma e \to Ze$ as a function of $y$ for $P = 90\%(0, -90\%)$ corresponding to the solid(dotted,dashed) curves. A $10^\circ$ angular cut has been imposed, and we assume that the photons are completely polarized.

Before continuing, let us use the expressions above to examine the $e^-$ beam polariza-tion dependence of the total SM cross section as well as the polarization asymmetry, $A_{DHG}$. (With NLC type detectors in mind we will make a $10^o$ angular cut in performing the integration above. We will also assume that the backscattered laser photons are 100% polarized.) Figs. 2a and 2b show the results of these considerations for three different beam polarizations as a function of $y = \sqrt{s}/M_Z$. The cross section maximum and polarization zero for all three cases occur at $y \simeq 1.27$ and $y \simeq 1.635$, respectively. For $P > (<)0$, the cross section is enhanced(suppressed) by a simple factor of $1 + PA_0$ with $A_0 \simeq 0.147$. Thus marginally higher statistics for total cross section determinations are obtained with left-handed beam polarization. The effect of $P \neq 0$ is much more significant in the case of $A_{DHG}$ and occurs due to the overall factor of $P + A_0$ which appears in front of the $\sigma_{pol}$ term. Here we see that when $P$ is non-zero and large, $A_{DHG}$ also has a significant magnitude (and is slightly bigger for the left-handed polarization case) almost everywhere but near the zero; this is quite fortuitous as the statistical error in $A_{DHG}$ is proportional to $\sqrt{1 - A_{DHG}^2}$ ! Clearly it is to our advantage to have highly polarized beams if we are to use $A_{DHG}$ as a probe for anomalous $\gamma ZZ$ couplings. From these considerations we will assume that the incoming $e^-$ beam is 90% left-handed polarized in our analysis below. This same assumption was made in our study of the $\gamma e \rightarrow W\nu$ process.

Let us first consider how the location of the polarization zero at $y_0 \simeq 1.635$ changes when $h_{3,4}^Z$ are non-zero. For simplicity, let us briefly ignore the energy dependence introduced from the form factors. A short calculation then leads one to the rather disappointing result

$$y_0 \simeq 1.63539 - 0.06530(h_3^Z)^2 - 0.07900(h_4^Z)^2 - 0.07550 h_3^Z h_4^Z , \qquad (8)$$

which tells us that the zero moves only very slightly, even for values of the anomalous couplings comparable to the present experimental limits, and that the zero's position has comparable sensitivity to non-zero values of either $h_3^Z$ or $h_4^Z$. This can be seen explicitly in Fig.3 where the largest effect occurs when both $h_3^Z$ and $h_4^Z$ have the same sign and can produce a coherent effect on the shift. From these considerations it is clear that a measurement of *only* the polarization zero's position will not be terribly useful in constraining the anomalous $\gamma ZZ$ couplings.

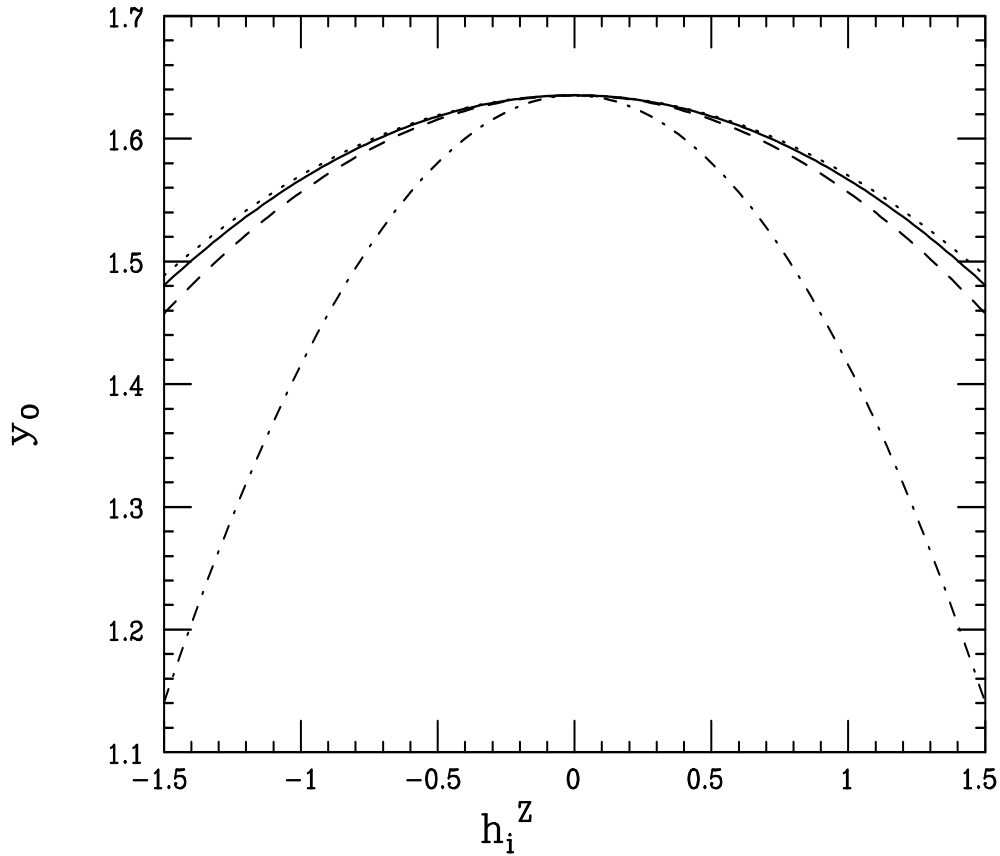

Figure 3: Position of the SM polarization asymmetry zero in $\gamma e \to Ze$ as a function of $h_{3,4}^Z$ subject to the assumptions discussed in the text. The dotted(dashed, dash-dotted, solid) curve corresponds to the case $h_4^Z = 0 (h_3^Z = 0,\ h_3^Z = h_4^Z,\ h_3^Z = -h_4^Z)$.

Of course we are not just interested in the zero's position but also in the energy dependence of the polarization asymmetry as a whole, as well as in the total cross section. To explore the effects on these two observables for small values of the anomalous $\gamma ZZ$ anomalous couplings including the contributions due to the form factors, let us assume $P = 90\%$ and take $\theta_0 = 10^o$ as above, together with a form factor scale of $\Lambda$=1.5 TeV. Fig. 4 then shows the resulting modifications in the total cross section and asymmetry. We see that the deviations in both observables are of comparable size and that most of the shift due to anomalous couplings occurs at large $y$ (i.e., large energies) as we might have expected. For the values of the couplings shown in the figures, we begin to see the deviations for $y$ values $\geq 3 - 4$. Of course to truly determine how sensitive these observables are to the anomalous couplings we turn to a Monte Carlo study along the line of our previous analysis[9]. We may anticipate from the these two figures, however, that both these observables will be far more sensitive to non-zero $h_4^Z$ than to non-zero $h_3^Z$. What about other observables? Two possibilities include the Z boson's angular distribution and polarization[10]. General arguments suggest that the angular distribution may not be overly sensitive to the existence of anomalous couplings unless the energies involved in the process are quite large. First, we note that in the backwards($z = -1$) direction, the $u$-channel pole greatly enhances the SM piece while no corresponding enhancement occurs in the anomalous moment contributions to the cross section. Second, if we go to the forward($z = 1$) direction where the SM contribution is well-behaved, the anomalous terms are seen to *vanish* due to the overall Bose symmetry of the coupling structure. The impact of the angular distribution on anomalous coupling constraints will be briefly discussed in the Appendix.

As a first pass at obtaining simultaneous constraints on the anomalous $\gamma ZZ$ couplings, let us consider $\gamma e$ collisions at a 500 GeV NLC. Due to the backscattered laser photon luminosity distribution, the $y$ range of interest here is approximately $1 \leq y \leq 4.6$. For simplicity, let us divide this range into 18 bins of width $\Delta y$=0.2 and assume a total luminosity of $54 fb^{-1}$ equally divided among these bins, i.e., $3 fb^{-1}$/bin. Next we follow the same procedure as in our earlier analysis of the $\gamma e \to W\nu$ process and generate Monte Carlo

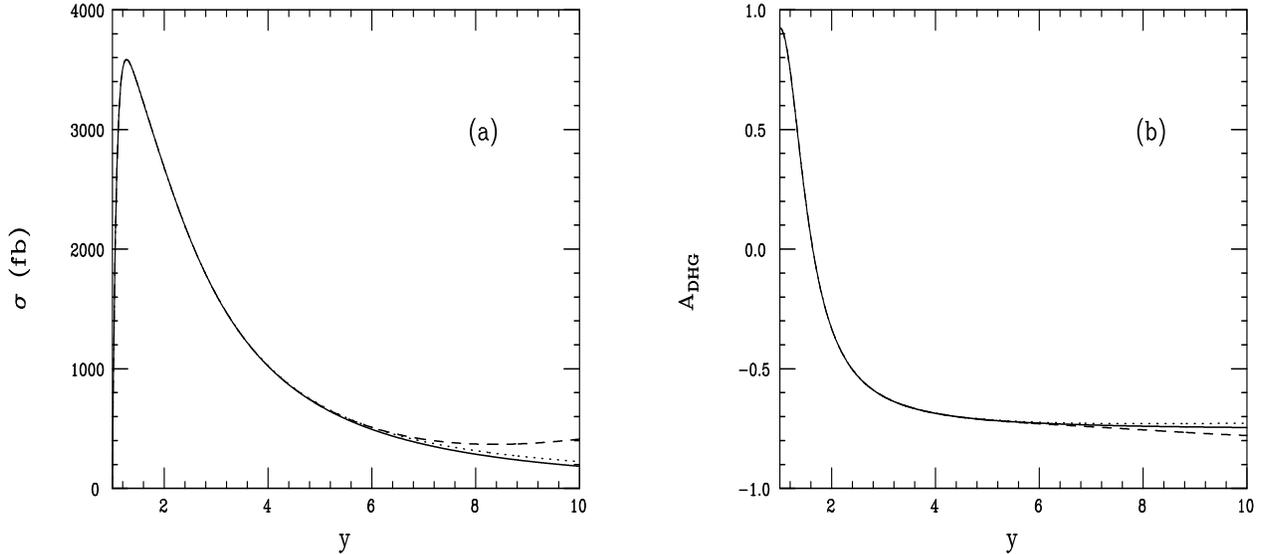

Figure 4: (a) Cross section and (b) polarization asymmetry in $\gamma e \to Ze$ as a function of $y$ for $P = 90\%$, $\theta_0 = 10°$, and $\Lambda = 1.5$ TeV. The solid(dotted,dashed) curves correspond to the SM($h_3^0=0.01$, $h_4^0=0.001$), respectively.

'data' for both the cross section as well as the asymmetry, assuming the validity of the SM, and try to fit the resulting distributions to the $h_{3,4}^0$-dependent functional forms given above. (We will include only statistical errors in performing this procedure.) Allowing both of the anomalous coupling parameters to be non-zero simultaneously we obtain the 95% CL region to the left of the solid curve shown in Fig. 5. As we expected, we obtain a much stronger constraint on $h_4^0$ than we do on $h_3^0$. If only one of the two anomalous couplings is non-zero we then obtain $|h_3^0| \leq 0.029$ and $|h_4^0| \leq 0.0054$ at 95% CL. These constraints lead to a somewhat smaller allowed region than can be obtained at a 2 TeV Tevatron with an integrated luminosity of $10 fb^{-1}$ and are about a factor of 3-4 worse than what is obtainable at the 14 TeV LHC with an integrated luminosity of $100 fb^{-1}$. We might expect that somewhat better limits might be obtainable by the optimization of our data binning and/or adjustment of our luminosity distribution as noted in our earlier work[9]. These expectations will be realized in the analysis which follows.

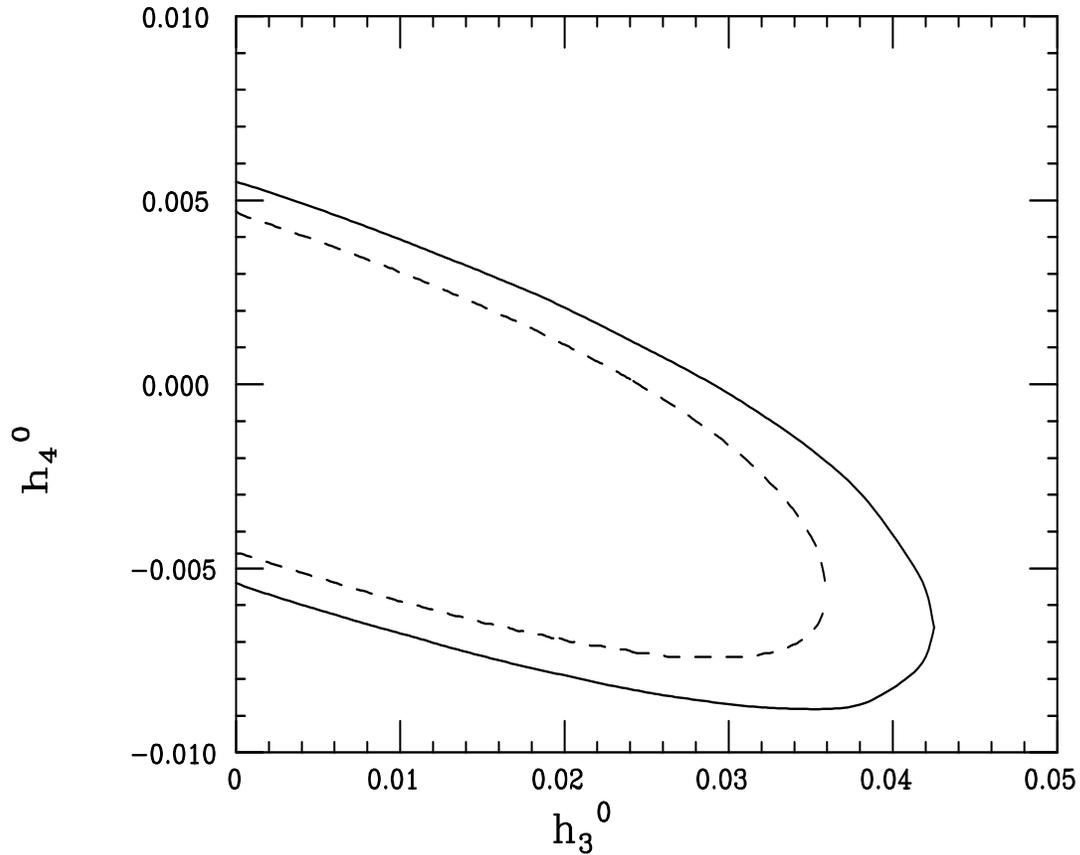

Figure 5: 95%$CL$ allowed region for the anomalous coupling parameters $h_3^0$ and $h_4^0$ from a combined fit to the energy dependencies of the total cross section and polarization asymmetry at a 500 GeV NLC assuming $P = 90\%$ and an integrated luminosity of $3(6) fb^{-1}$/bin corresponding to the solid (dashed) curve. 18 bins of width $\Delta y$=0.2 were chosen to cover the $y$ range $1 \leq y \leq 4.6$. The corresponding bounds for negative values of $h_3^Z$ are obtainable by remembering the invariance of the polarization dependent cross section under the reflection $h_{3,4}^0 \to -h_{3,4}^0$.

To examine the influence of limited statistics on these results, we have repeated our analysis above after a simple doubling of the integrated luminosity per bin and keeping the collider energy fixed, *i.e.*, we now assume $6fb^{-1}$/bin. The result of this analysis is the dashed curve in Fig. 5. While our 95% CL region does shrink, the decrease is not large which implies that it is probably more important to go to higher energies than higher luminosities if one wishes to improve the constraints on the anomalous $\gamma ZZ$ couplings.

We now turn our attention to a 1 TeV NLC with which both larger values of $y$ become accessible and higher integrated luminosities are available. As a first pass in this case, we consider taking 42 bins of width $\Delta y$=0.2 thus covering the range $1 \leq y \leq 9.4$. We will initially assume a uniform luminosity distribution of $4fb^{-1}$/bin for a total integrated luminosity of $168fb^{-1}$ similar to the procedure above. As before we have not tried to optimize these particular choices. Generating 'data' as before assuming the validity of the SM and performing the fit we obtain the 95% CL range in Fig. 6 assuming both $h_3^0$ and $h_4^0$ are simultaneously non-zero. (In the case that only one of these is non-zero, we obtain the very strict constraints $|h_3^0| \leq 0.0066$ and $|h_4^0| \leq 0.00033$.) These limits are reasonably comparable to those obtainable at the 14 TeV LHC with a luminosity of $100fb^{-1}$ and about a factor of 2-3 better than that obtained at the 500 GeV NLC we discussed above. In order to examine the statistical limitations of these results let us again consider the influence of doubling the integrated luminosity while keeping the collider energy fixed. This results in a somewhat shrunken allowed region corresponding to that which is within the dashed curve in Fig. 6. Although this is an improvement, the gain is not that large.

The deviations in both $\sigma$ and $A_{DHG}$ due the presence of the $\gamma ZZ$ anomalous couplings shown in Fig. 4 remind us that most of the sensitivity arises at large values of $y$. To this end we reconsider the 1 TeV case with a total integrated luminosity of $168fb^{-1}$, but instead of distributing the luminosity equally, we now only distribute it in a flat distribution over the last 21 bins. This means that these higher energy bins now receive $8fb^{-1}$/bin instead of $4fb^{-1}$/bin. The result of this procedure is the dotted curve in Fig. 6, from which we learn that almost all the improvement obtained from doubling the luminosity in our earlier study arose from the high-$y$ end of the fit to the data. This implies that for a fixed energy and total integrated luminosity we should preferentially distribute the luminosity into the highest energy bins. Lastly, pushing this idea a bit further, we place $16.8fb^{-1}$ of luminosity in each of the last 10 bins and we find the result given by the dash-dotted curve in Fig. 6. These limits are quite competitive with those from the LHC, but carve out a different region in the $h_3^0 - h_4^0$ parameter space implying that the results from both machines will be complementary. If only one of the anomalous couplings can differ from zero at a time we now obtain the corresponding 95% CL constraints $|h_3^0| \leq 0.0045$ and $|h_4^0| \leq 0.00023$, respectively. We thus see quite explicitly that by putting all the luminosity in the highest energy bins we gain substantially in reducing the size of the allowed region of the anomalous couplings for a fixed total integrated luminosity. The moral of this study is that in obtaining limits on $\gamma ZZ$ anomalous couplings higher energy wins over higher statistics. In particular, in going from the situation where all bins receive the same luminosity to the case where all the luminosity goes into the last 10 bins (with fixed total luminosity), we gain about a factor of two in the combined constraints on $h_3^0$ and $h_4^0$.

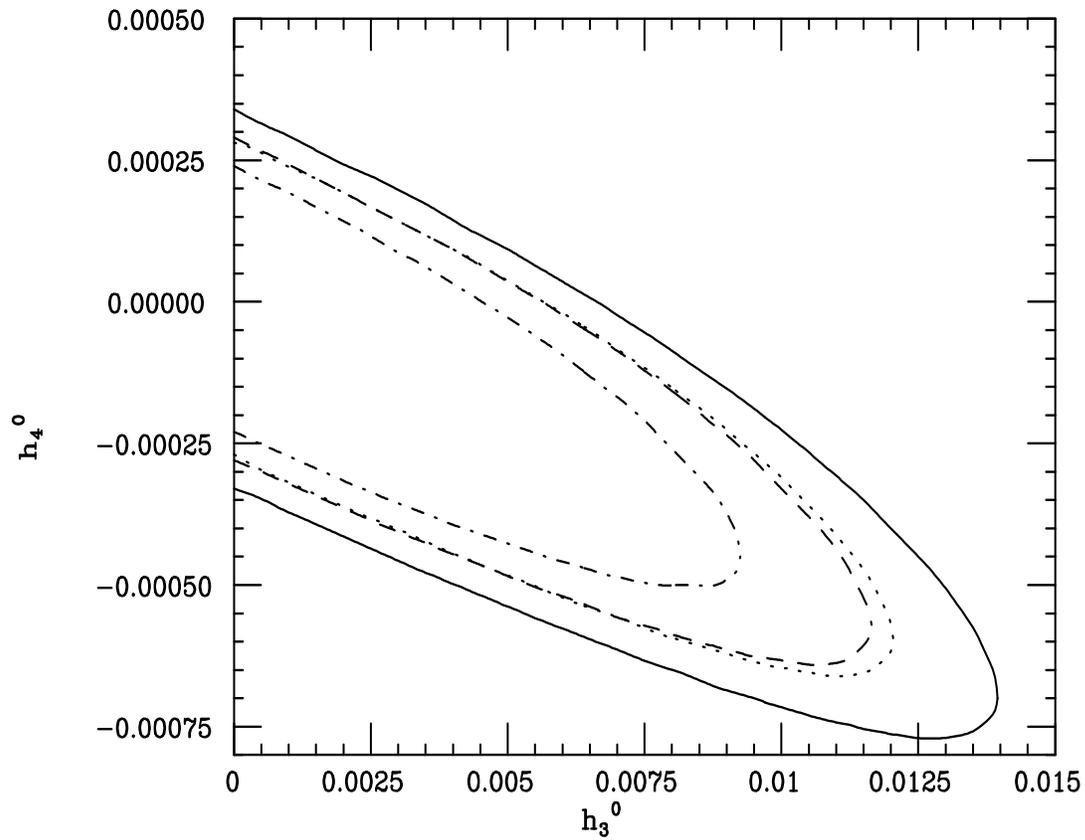

Figure 6: Same as Fig. 5 but for a 1 TeV NLC. The solid(dashed) curve corresponds to a luminosity of $4(8)fb^{-1}$/bin for 42 bins of width $\Delta y$=0.2 which covered the range $1 \leq y \leq 9.4$. The dotted curve corresponds to a luminosity of $8fb^{-1}$/bin but only for the last 21 bins. The dash-dotted curve corresponds to the case of $16.8fb^{-1}$/bin in only the last 10 bins.

## 3  Discussion and Conclusions

While Z pole measurements have led to a remarkable improvement in our knowledge about the couplings of gauge bosons to fermion pairs, we are still rather ignorant about the trilinear gauge couplings from the point of view of direct experimental probes. To truly test the SM, all such trilinear couplings must be examined experimentally with very high precision which future hadron and $e^+e^-$ colliders will allow us to do.

In this paper we focussed on the capability of the NLC in the $\gamma e$ collision mode to probe the CP-conserving $\gamma ZZ$ anomalous couplings via the process $\gamma e \to Ze$. In particular, we examined the energy dependencies of both the total cross section as well as the polarization asymmetry as constraints on these anomalous couplings. As in our previous analysis of the $\gamma e \to W\nu$ process, we found this polarization asymmetry to be quite sensitive to the presence of such couplings. Unlike the previous case, however, the position of the asymmetry zero as a function of energy was itself found to be rather insensitive to the anomalous couplings. For the SM case the position of the zero occurs at much lower energies than does the corresponding one in the $\gamma e \to W\nu$ process, lying in the 130-150 GeV range, depending upon the angular acceptance cuts applied to identify the $\gamma e \to Ze$ process.

The results of our Monte Carlo study are as follows:

($i$) The cross section and polarization asymmetries for $\gamma e \to Ze$ yield comparable constraints on the the $\gamma ZZ$ anomalous couplings.

($ii$) For either a 500 GeV or 1 TeV NLC the constraints imposed by considering jointly the energy dependencies of the total cross section and polarization asymmetry for the $\gamma e \to Ze$ process are tighter on $h_4^0$ than on $h_3^0$.

($iii$) For a fixed $e^+e^-$ center of mass energy and total $\gamma e$ integrated luminosity stronger constraints are obtainable if the luminosity is concentrated in the highest energy bins. This improvement can be superior to a doubling of the total luminosity and keeping a flat distribution over the bins flat.

($iv$) The bounds we obtained on the $h_{3,4}^0$ parameters at a 500 GeV NLC were found to be superior to what could be obtained at the Tevatron and only a factor of 3-4 worse than that achievable at the LHC. Increasing the NLC energy to 1 TeV and distributing the luminosity to the highest energy bins, we found constraints comparable to the LHC. However, the allowed regions in the $h_3^0 - h_4^0$ plane obtained at the hadron colliders and the NLC have different shapes implying the complementarity of the two sets of machines in constraining anomalous $\gamma ZZ$ couplings. This is clearly shown by Fig. 8 and the discussion in the Appendix.

($v$) We remind the reader that there are additional observables available at the NLC that we have not used in this analysis that can be called upon to further constrain these anomalous couplings using the $\gamma e \to Ze$ process—two examples of which are the $Z$ boson's angular distribution and polarization. A complete analysis of the $\gamma e \to Ze$ process at the NLC will, of course include all of these quantities[10]. Incorporating both these observables into the analysis will most likely yield constraints superior to the LHC from this single process. The analysis presented in the Appendix justifies these expectations as is shown in Fig. 8. In addition, the crossed process $e^+e^- \to Z\gamma$ can also be used to further restrict (or discover!) the $\gamma ZZ$ anomalous couplings.

The search for the influence of anomalous gauge couplings at colliders is just beginning.

# 4  Acknowledgements

The author would like to thank J.L. Hewett, S.J. Brodsky and I. Schmidt for discussions related to this work. He would also like to thank the members of the particle theory group of the Technion for their hospitality and use of their computing facilities while this work was in progress.

**Appendix**

The purpose of this Appendix is to briefly examine the additional constraints imposed on the CP-even anomalous $\gamma ZZ$ couplings $h_3^0$ and $h_4^0$ which arise through an analysis of the $\gamma e \to Ze$ angular distribution of the unpolarized cross section. As mentioned in the text, the contributions to this distribution from the anomalous couplings essentially vanish at either end point so that most of the sensitivity must arise from the central region. To this end, we show in Fig. 7 the SM distribution, as well as the corresponding distributions when the anomalous coupling terms are present, for three different values of $\sqrt{s_{\gamma e}}$. In all cases $P = 90\%$ and a scale $\Lambda = 1.5$ TeV are assumed as in the main body of this paper. Note that at lower energies the anomalous coupling contributions are almost invisible but increase rapidly in significance as $\sqrt{s_{\gamma e}}$ is increased.

In order to make use of the angular distribution, we return to our Monte Carlo study. In particular, let us re-examine the data set corresponding to the 1 TeV NLC where the luminosity was equally distributed over the last 10 $\Delta y$ bins with $16.8 fb^{-1}$/bin and deconvolute the $-z_0 \leq z \leq z_0$ integration. (Recall that $z_0 = cos(10^o)$ in this analysis.) Now we place the data in 10 bins of $\Delta z = 0.2$ (except for the two end bins which have a width of $0.2 - (1 - z_0)$), and integrate over the $y$ range $7.4 \leq y \leq 9.4$. Next, we combine the fit of this data to the $y$-integrated, but $h_{3,4}^0$-dependent, distribution for the unpolarized differential cross section (obtainable from Eqs. 5-7), together with the corresponding fit to the energy dependence of the total cross section and $A_{DHG}$ as performed in Section 3. The result of this *simultaneous* fit is shown by the 95% CL region enclosed by the solid curve in Fig. 8; the corresponding one-dimensional fits give $|h_3^0| \leq 0.0033$ and $|h_4^0| \leq 0.00020$, respectively. In either case, the fit to the angular distribution has provided additional constraints on the $\gamma ZZ$ anomalous couplings. For the LHC, the corresponding bounds on the anomalous couplings are shown by the dotted curve[8] demonstrating the complementarity of the two sets of measurements. We note that the overlap region obtained by combining the NLC and LHC results is very small.

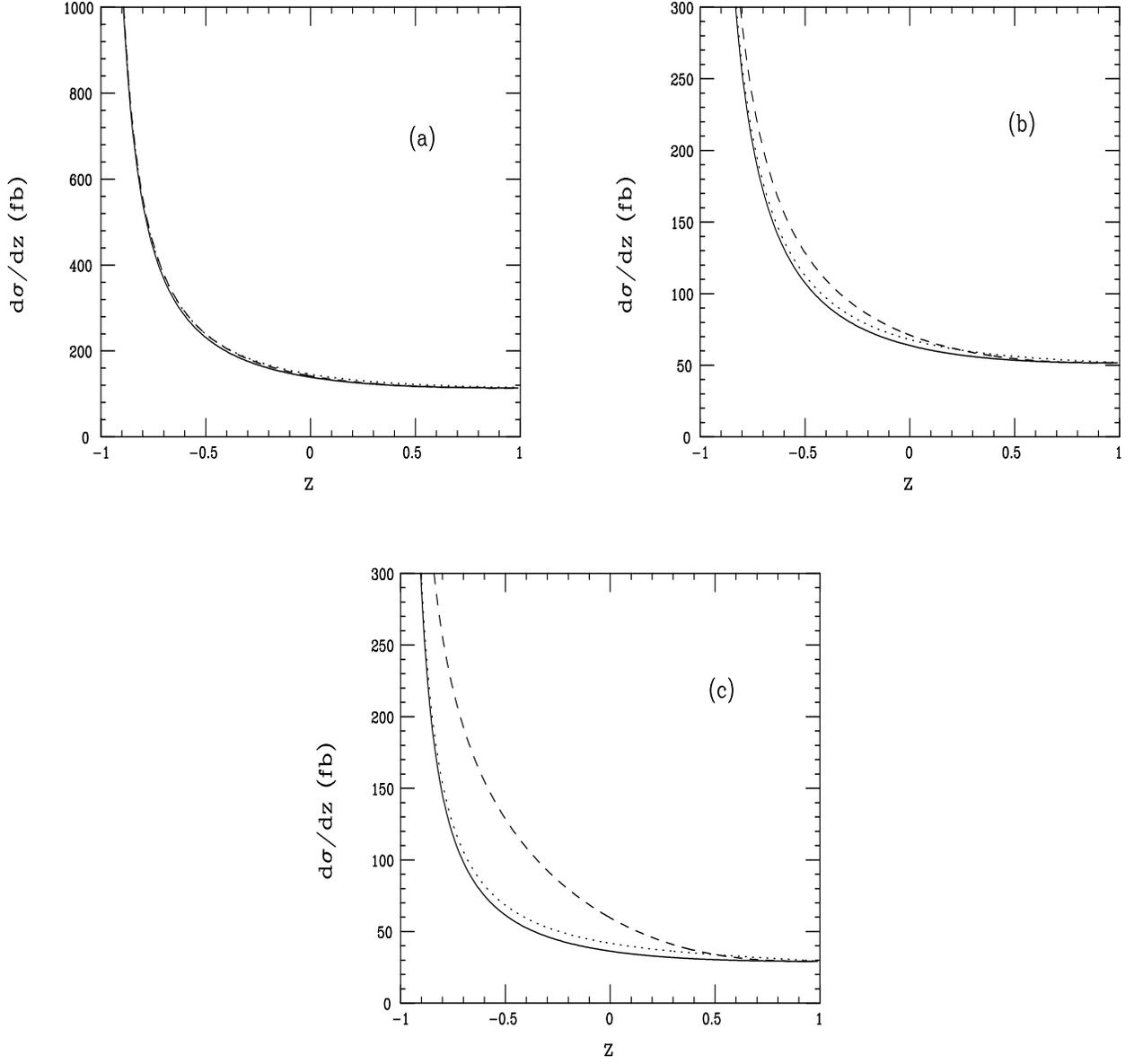

Figure 7: Angular distribution for $\gamma e \to Ze$ assuming $P = 90\%$ and $\Lambda = 1.5$ TeV. The SM is the solid curve in all three figures. In (a)[(b), (c)], the $\gamma e$ center of mass energy is 500[750, 1000] GeV. In (a), the dotted(dashed) curve corresponds to $h_3^0 = 0.01 (h_4^0 = 0.001)$, while in both (b) and (c) the value is taken to be 0.005(0.0005).

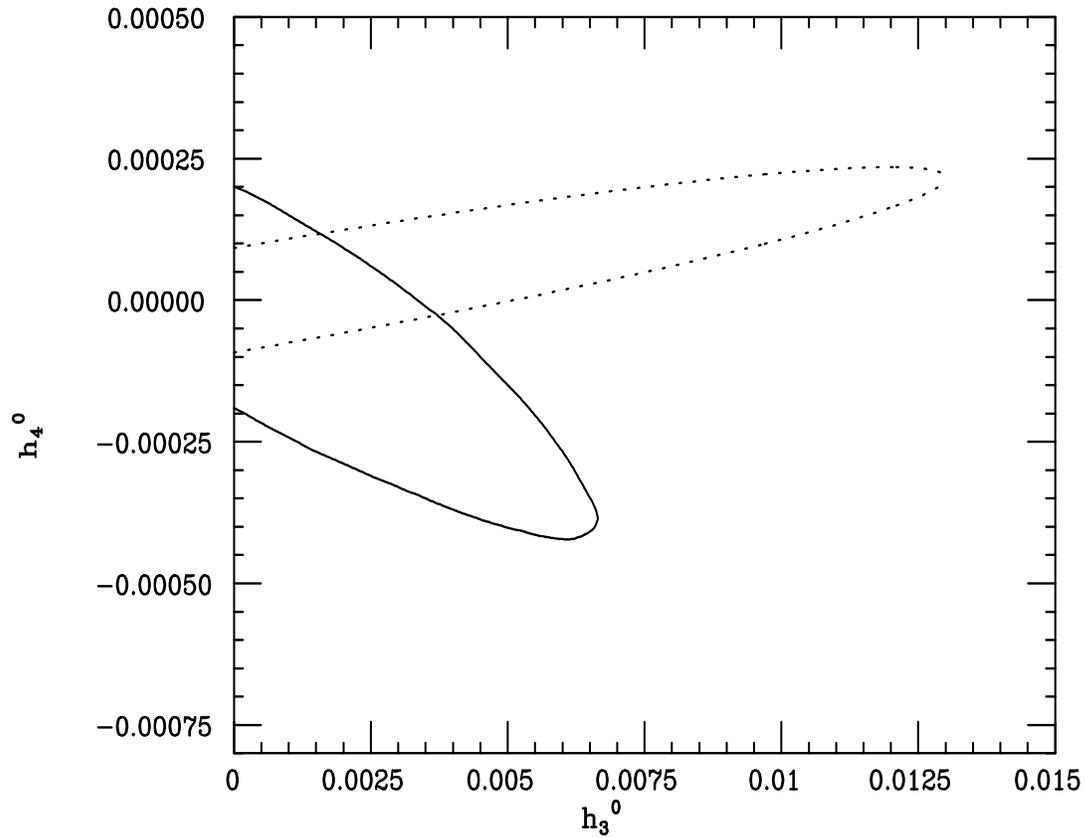

Figure 8: Same as dash-dotted curve in Fig. 6, but now including the fit to the angular distribution obtained from the highest 10 bins in energy. The corresponding result for the 14 TeV LHC is the dotted curve.